\documentclass[varenna]{cimento}
\usepackage{latexsym}
\usepackage{amsmath,amssymb}
\usepackage[retainorgcmds]{IEEEtrantools}
\usepackage[T1]{fontenc}
\usepackage{graphicx}
\usepackage{sidecap}
\title{New quarks: exotic versus strong}
\author{B.~Holdom}
\institute{Dept. of Physics, University of Toronto, Toronto ON, Canada}
\PACSes{\PACSit{14.65.Jk}{Fourth generation quarks}}
\begin{document}

\maketitle

\begin{abstract}
The new quarks of a fourth family are being pushed into the strongly interacting regime due to the lower limits on their masses. The theoretical basis and experimental implications of such quarks are compared with exotic quarks.
\end{abstract}

New quarks can come in one of three varieties: 1) a fourth family with Higgs, 2) a ``strong'' fourth family without Higgs, 3) ``exotic'' quarks (having nonstandard quantum numbers).

Adding a fourth family to the standard model is perhaps its simplest extension, and the SM4 was receiving attention recently after decades of neglect. In this model the Higgs couples to the heavy quarks in the standard way, and this leads to a substantial enhancement of the $gg\rightarrow H$ cross section \cite{Arik:2005ed}. This in turn leads to dramatic exclusion limits on the Higgs assuming SM4. From this it is sometimes concluded that the fourth family is in `deep trouble' \cite{Peskin:2011zz}. The implicit assumption being made is that the standard Higgs exists.

To be more precise the data indicates that the fourth family and the standard Higgs cannot both exist. Possibility (1) above is being excluded. Turning it around, if the fourth family is found then it is the standard Higgs that is in deep trouble. This raises the stakes for the fourth family search, and so given this it is rather puzzling that this search has not attracted more attention by the experimentalists. Nevertheless, direct searches are occurring and the lower limits on the fourth family quark masses have increased. Currently these limits are in the 450 to 500 GeV range.

In the context of SM4 it is useful to review the theoretical implications of a fourth family above 500 GeV. First the large Yukawa couping $y_{q'}$ has a large impact on the running of the quartic Higgs coupling since $\mu d\lambda/d\mu\propto \lambda y_{q'}^2-y_{q'}^4+...$ \cite{Kribs:2007nz}. This means that the allowed range of the Higgs mass is greatly diminished, if it exists at all, for $m_{q'}>500$ GeV. The Yukawa coupling $y_{q'}(\mu)$ also more quickly runs into trouble. Probably most dramatic is the direct contribution to the Higgs mass $\delta m_h^2\approx ({m_{q'}}/{\mbox{400 GeV}})^2\Lambda^2$. Here $\Lambda$ represents the scale of the physics that cuts off the quadratic divergence. Thus the Higgs mass is driven up to the cutoff even as $m_{q'}$ moves above 400 GeV. These theoretical considerations indicate that the fourth family with large mass cannot co-exist with the standard Higgs. And it is this result that the experimental results are confirming.

Thus far the heavy quark searches at the LHC have employed quite basic strategies. The $t'$ has been searched for in the $\ell+\mbox{jets}$ mode ($H_T$ and $M_{\rm recon}$ distributions) and in the dilepton mode ($M_{b\ell}$ distributions). The $b'$ search employs same-sign leptons. Are these actually the best strategies for large quark masses? For example, 600 GeV masses would only have produced a few same-sign lepton events thus far. We also note that the current best limits on heavy quark masses are coming from CMS only. ATLAS is strangely quiet about heavy quarks; perhaps they are exploring more promising strategies.

Let us consider the kinematics of a heavy quark search. For $m_{t'}>500$ GeV the process $t'\overline{t'}\rightarrow b\overline{b}WW$ will produce $W$'s that are typically both boosted and isolated. The jets from their hadronic decay will often merge to produce a single $W$-jet. Meanwhile the background can be suppressed with an $H_T$ cut, such as $H_T\gtrsim2m_{t'}$. In this case the dominant $t\overline{t}$ background will tend to look like the production of boosted tops, in which case the $W$'s are not isolated from their associated $b$'s. This different kinematics of signal and background should be exploited \cite{Holdom:2007nw}.

As far as I know a direct search for isolated $W$-jets has not been undertaken. The more difficult problem of extracting $W$-jets from boosted tops has been shown to be feasible through the use of a large cone size and jet pruning techniques \cite{CMS6}. A search for isolated $W$-jets would presumably be simpler, and in fact we would want a low efficiency for identifying $W$-jets from boosted tops since the latter is a feature of the background. And finally we note that both $t'$ and $b'$ production would contribute to the isolated $W$-jet signal.

Details of a search strategy for isolated $W$-jets can be found in \cite{Holdom:2010za} and we choose not to repeat it here. The use of $W$-jets in a simple reconstruction of the $t'$ mass in the $\ell+\mbox{jets}$ decay mode of $t'\overline{t}'$ was also considered, and here we would like to describe an improvement of this method. It makes full use of the boosted characteristics of both the hadronic and leptonic $W$'s.

The reconstruction uses an anti-$k_T$ jet finder with $R=.8$ and involves three objects: 1) a $W$-jet ($p_T>100$ GeV and mass within 12 GeV of $M_W$), 2) a leptonic $W$, and 3) either a $b$-jet ($p_T>50$ GeV) or a non-$W$-jet ($p_T>100$ GeV). The leptonic $W$ is assumed to be sufficiently boosted so that the lepton and neutrino are close to the same direction. Thus the missing transverse energy can be used along with an isolated lepton ($p_T>20$ GeV) to reconstruct the leptonic $W$ momentum. Now the procedure is to take each object of the third type and pair it with the $W$ (hadronic or leptonic) that gives the largest invariant mass. Repeat this for all possible selections of the three objects in each event and then histogram these invariant masses.
\begin{figure}
\begin{center}\includegraphics[scale=0.26]{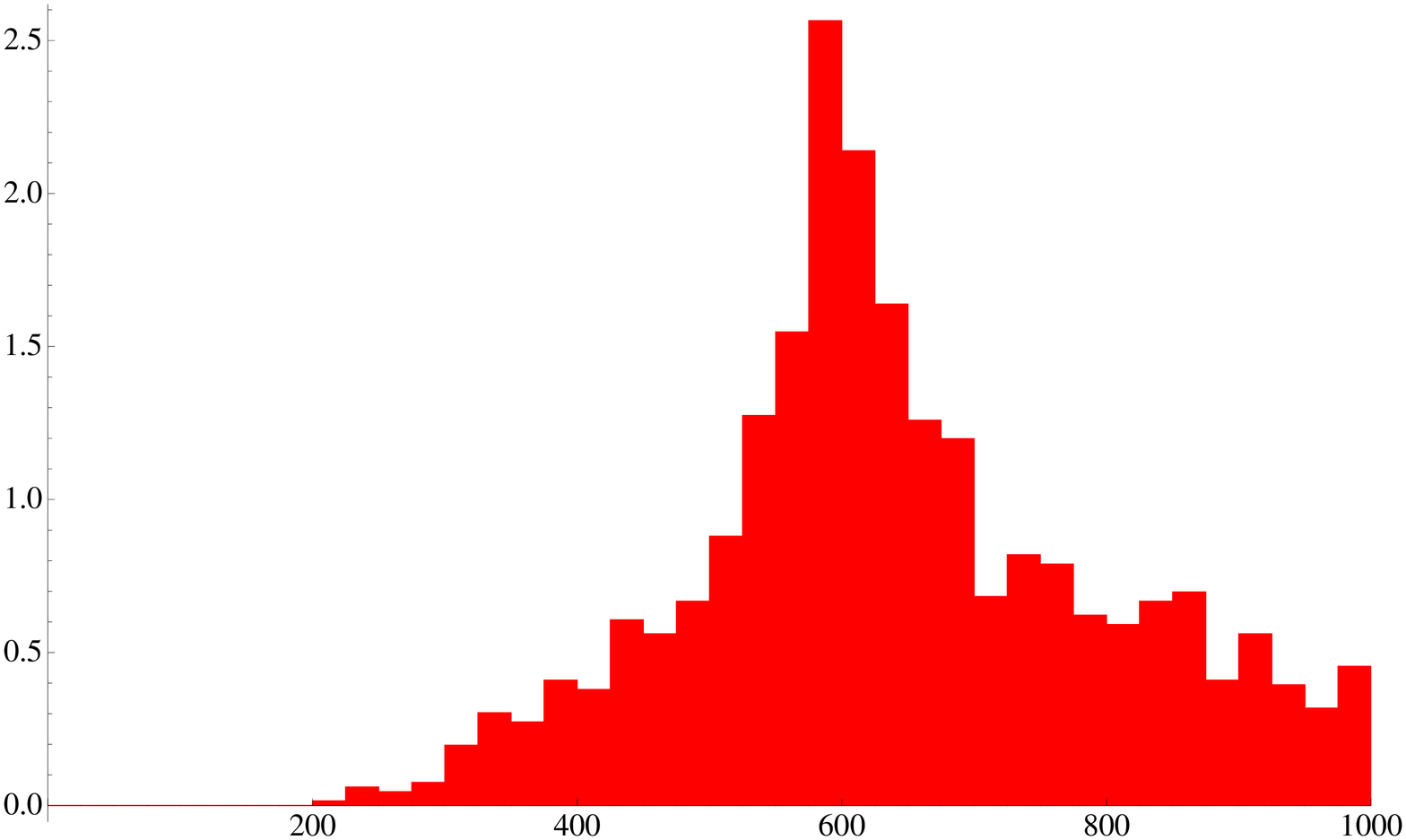}\quad\includegraphics[scale=0.26]{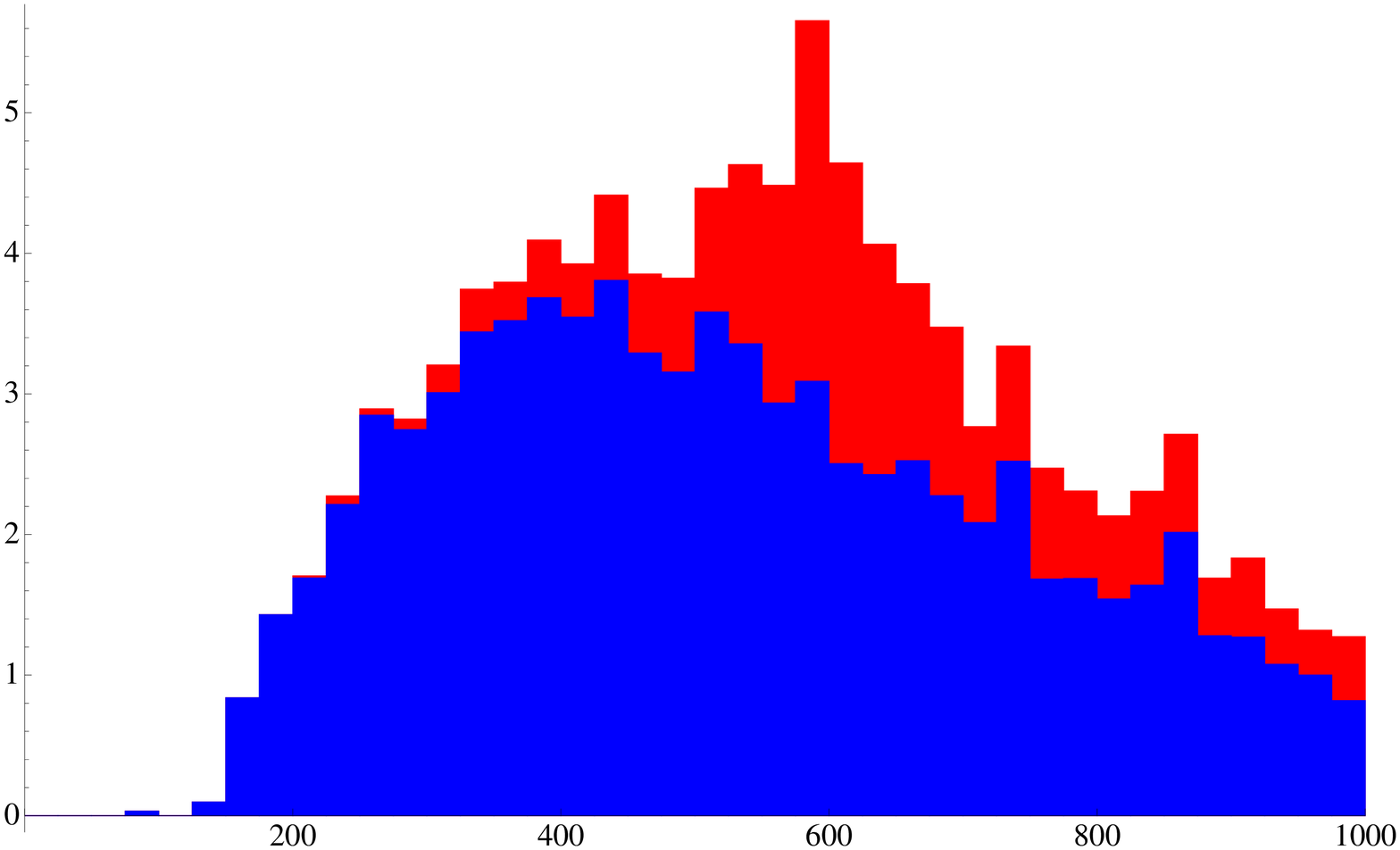}\end{center}
\caption{Signal (left) and signal plus $t\overline{t}$ background (right) for 2.5 fm$^{-1}$ ($\sqrt{s}=7$ TeV).}
\label{f1}\vspace{-5ex}\end{figure}

If each event is required to have a $b$-jet then $t\overline{t}$ will be the dominant background. We also impose the $H_T>1100$ GeV cut. We present the results in Fig.~\ref{f1} for $m_{t'}=600$ GeV. We find a signal that is quite prominent over a background which is quite featureless at the mass of the heavy quark. It appears that this method provides a more powerful search strategy than the ones currently being employed for the $t'$ search.

We now turn to exotic quarks. The interest here is in ``vector-like'' fields such that the left and right-handed fields transform the same under $SU(2)_L\times U(1)$. Then the quark masses are independent of electroweak symmetry breaking which in turn provides much freedom for model building. Ref.~\cite{delAguila:2000rc} nicely summarizes the possible exotic quarks transforming as singlets, doublets or triplets under $SU(2)_L$. They are chosen such that they can mix with standard quarks through Yukawa terms.

There was 	a period of time where exotic quarks were considered the only `game in town' for new quarks. For example ``a fourth generation of chiral fermions is excluded at 99\% C.L. by the present limits on the $S$ parameter'' \cite{delAguila:2000rc}. But such claims are now realized to be a bit premature \cite{Holdom:2006mr,Kribs:2007nz}.

Various features of the standard model, such as the CKM description, no tree-level FCNC, no Higgs mediated FCNC and no right-handed charged currents, among others, are broken when exotic quarks are added \cite{delAguila:2000rc}. To avoid problems the exotic quarks typically mix mainly just with the third family. Thus in various models of exotic quarks one often finds that the third family is special in some way. This is evident for topcolor with a see-saw mechanism for the top mass, warped extra dimension with excited states of third family quarks, little Higgs with top quark partners to cancel Higgs mass contributions, and the substantial mixing with a composite top in composite Higgs models. The masses of exotic quarks in these models span a large range, with no real preference for masses near the low end of the allowed range.

Exotic quarks have exotic decays. For example for the $SU(2)_L$ singlets $U_L$ and $U_R$ the mixing occurs through the term $Y\overline{q}_LU_R\tilde{\phi}+hc$. Since $\phi$ describes both the Higgs and Goldstone fields this term produces the decays $U\rightarrow Wb,\;Zt, Ht$. For $SU(2)_L$ doublet quarks $Q_L$ and $Q_R$ with $Q=(U,D)$ the terms are ${Y_t}\overline{Q}_Lt_R\tilde{\phi}+{Y_b}\overline{Q}_Lb_R\phi+hc$. Here $U\rightarrow Wb,\;Zt, Ht$ and $D\rightarrow{Wt}{,Zb, Hb}$. In either case the proportions of $W:Z:H$ produced are $\approx1/2:1/4:1/4$. This means that exotic quarks have quite a firm prediction for  the production of $Z$'s through the process $Q\overline{Q}\rightarrow Z+X$. This is in contrast to the single production of exotic quarks which depends on a very model dependent mixing parameter.

Once a heavy quark is found it will be crucial to decide whether it is exotic or part of a fourth family. It was seen in \cite{Holdom:2010za} that a search based on the two lepton decay of $Z$'s from exotic quarks has a similar sensitivity to the same-sign lepton search for fourth family quarks. This is reflected by the fact that the current searches for $Q\overline{Q}\rightarrow Z+X$ are producing similar limits on the quark masses as the same-sign lepton limits.

From our discussion of exotic quarks we see that they are usually associated with attempts to protect the Higgs. In fact they can be viewed as part of a massive theoretical effort to keep electroweak symmetry breaking perturbative. The proposed models often have strong interactions and/or other complications occurring at somewhat higher energies instead. If the LHC was to discover strong interactions at the TeV scale then this would diminish the motivation for exotic quarks.

Vector meson $\rho$-like resonances are usually thought to be a generic feature of new strong interactions (technicolor, Higgsless models etc.). But this is not the only possibility. The new massive states to be seen first may well be fermionic. This can occur when the strong interactions produce a dynamical mass for the fermions without confining these fermions. The physical states can then correspond to the elementary fermions.

This then returns us to the fourth family, since this is obviously the first guess for what a new set of fermions should look like. In this case the dynamical quark masses are a manifestation of dynamical electroweak symmetry breaking. The quark masses are bounded from above by unitarity and also by their connection with the $W$ and $Z$ masses. The upper bound on the mass, around 600 or perhaps even 700 GeV, ensures that the LHC will either find these quarks or rule them out, and so this makes the case for their search even more compelling.

To emphasize the connection with electroweak symmetry breaking, we note that the Goldstone bosons are fluctuations in the EWSB order parameter. But we know that the Goldstone bosons couple strongly to the heavy quarks given that their masses are $\gtrsim 500$ GeV. It is then to be expected that condensates of fourth family fermions will be the EWSB order parameters. Thus once again we see that the Higgs does not belong in this picture \cite{Holdom:2006mr}.

We can mention very briefly some of the underpinnings of this picture. The formation of quark condensates without confinement takes us back 50 years to the NJL model. The 4-fermion operators that formed the basis of that model can be generated by some new strong \textit{broken} gauge interaction. We emphasize that there is no need and no reason to expect fine tuning in the NJL-like interaction strength, and in this case there is no light Higgs-like composite scalar.

The broken gauge interaction can be a remnant of a broken flavor gauge interaction. The breakdown of this larger flavor gauge symmetry occurs mostly at a higher scale and gives rise to a wide variety of 4-fermion operators. Some of these can connect different families and have the effect of feeding mass down from heavy to light, so that $\overline{\Psi}\Psi\overline{\psi}\psi/{\Lambda^2}$ operators replace Yukawa couplings. This is enough to see that a heavy fourth family not only teaches us something new about EWSB, but it may also be the beginning of our understanding of flavor.

To sum up, I think it is fair to say that theorists are attracted to the ``exotic'' while avoiding the ``strong''. This helps to explain why the fourth family has been and still is overlooked and under appreciated. Another reason is that rather than helping to protect the Higgs, it replaces it. Thus will the fourth family be the David to the Higg's Goliath?

\acknowledgments
This work was supported in part by the Natural Sciences and Engineering Research Council of Canada.

\end{document}